\documentclass[conference,a4paper]{IEEEtran}

\usepackage[T1]{fontenc}
\usepackage[utf8]{inputenc}
\usepackage{lmodern}

\usepackage{amsmath,amssymb,amsfonts}

\usepackage{graphicx}
\usepackage{xcolor}
\usepackage{adjustbox}
\usepackage{wrapfig}
\usepackage{float}
\usepackage{tikz}

\usepackage{tabularx}
\usepackage{longtable}
\usepackage{multirow}
\usepackage{multicol}

\usepackage{listings}
\usepackage[inline]{enumitem}

\usepackage[nolist]{acronym}
\usepackage{csquotes}
\usepackage{comment}
\usepackage{bbding}   

\usepackage[hidelinks]{hyperref}

\usepackage[caption=false,font=footnotesize]{subfig}

\usepackage[nameinlink,noabbrev]{cleveref}
\crefname{subfigure}{Fig.}{Figs.}
\Crefname{subfigure}{Fig.}{Figs.}

\title{From ECU to VSOC: UDS Security Monitoring Strategies%
\thanks{Preprint (arXiv). Accepted to SECURWARE 2025, The Nineteenth International Conference on Emerging Security Information, Systems and Technologies, Barcelona, Spain. Reference: \url{https://www.thinkmind.org/library/SECURWARE/SECURWARE_2025/securware_2025_1_70_30030.html}}}

\author{%
\IEEEauthorblockN{Ali Recai Yekta\IEEEauthorrefmark{1}, Nicolas Loza\IEEEauthorrefmark{2}, Jens Gramm\IEEEauthorrefmark{2}, Michael Peter Schneider\IEEEauthorrefmark{2}, Stefan Katzenbeisser\IEEEauthorrefmark{3}}
\IEEEauthorblockA{\IEEEauthorrefmark{1}Yekta IT GmbH, Dortmund, Germany}
\IEEEauthorblockA{\IEEEauthorrefmark{2}ETAS GmbH, Stuttgart, Germany}
\IEEEauthorblockA{\IEEEauthorrefmark{3}University of Passau, Passau, Germany}
\IEEEauthorblockA{ali@yekta-it.de, \{nicolas.loza \,|\, jens.gramm \,|\, michaelpeter.schneider\}@etas.com, stefan.katzenbeisser@uni-passau.de}
}

\begin{document}
\maketitle

\begin{abstract}

\begin{abstract}
Increasing complexity and connectivity of mo\-dern vehicles have heightened their vulnerability to cyberattacks. This paper addresses security challenges associated with the Unified Diagnostic Services (UDS) protocol, a critical communication framework for vehicle diagnostics in the automotive industry. We present security monitoring strategies for the UDS protocol that leverage in-vehicle logging and remote analysis through a Vehicle Security Operations Center (VSOC). Our approach involves specifying security event logging requirements, contextual data collection, and the development of detection strategies aimed at identifying UDS attack scenarios. By applying these strategies to a comprehensive taxonomy of UDS attack techniques, we demonstrate that our detection methods cover a wide range of potential attack vectors. Furthermore, we assess the adequacy of current AUTOSAR standardized security events in supporting UDS attack detection, identifying gaps in the current standard. This work enhances the understanding of vehicle security monitoring and provides an example for developing robust cybersecurity measures in automotive communication protocols.
\end{abstract}


\end{abstract}

\begin{IEEEkeywords}
Automotive Networks; Automotive Security; UDS; Security Monitoring; VSOC; UN R155; IDS
\end{IEEEkeywords}


\begin{acronym}[xxxxxx]

\acro{as}[AS]{Attack Scenarios}
\acro{aws}[AWS]{Amazon Web Services}
\acro{secops}[SecOps]{Security Operations}
\acro{icmp}[ICMP]{Internet Control Message Protocol}
\acro{someip}[SOME/IP]{Scalable Service-Oriented Middleware over IP}
\acro{dds}[DDS]{Data Distribution Service}

\acro{misp}[MISP]{Malware Information Sharing Platform}
\acro{mqtt}[MQTT]{Message Queuing Telemetry Transport}
\acro{poc}[PoC]{Proof-of-Concept}
\acro{mti}[MTI]{Mobility Threat Intelligence}
\acro{adas}[ADAS]{Advanced Driver Assistance System}
\acro{ai}[AI]{Artificial Intelligence}
\acro{api}[API]{Application Programming Interface}
\acro{asn1}[ASN.1]{Abstract Syntax Notation One}
\acro{atl}[aTL]{advanced Train Lab}
\acro{ato}[ATO]{Automatic Train Operation}
\acro{atp}[ATP]{Automatic Train Protection}
\acro{autosar}[AUTOSAR]{Automotive Open System Architecture}
\acro{ber}[BER]{Basic Encoding Rule}
\acro{cv2x}[C\=/V2X]{Cellular Vehicle-to-Everything}
\acro{dos}[DoS]{Denial of Service}
\acro{dsrc}[DSRC]{Dedicated Short-Range Communication}
\acro{ecu}[ECU]{Electronic Control Unit}
\acro{ee}[E/E]{Electrical/Electronic}
\acro{etcs}[ETCS]{European Train Control System}
\acro{fota}[FOTA]{Firmware-over-the-air}
\acro{frmcs}[FRMCS]{Future Railway Mobile Communication System}
\acro{gdpr}[GDPR]{General Data Protection Regulation}
\acro{gvm}[GVM]{Generic Vehicle Model}
\acro{hids}[HIDS]{Host IDS}
\acro{http}[HTTP]{Hypertext Transfer Protocol}
\acro{ics}[ICS]{Industrial Control System}
\acro{ids}[IDS]{Intrusion Detection System}
\acro{ioc}[IoC]{Indicator of Compromise}
\acroplural{ioc}[IoC]{Indicators of Compromise}
\acro{isac}[ISAC]{Information Sharing and Analysis Center}
\acro{it}[IT]{Information Technology}
\acro{json}[JSON]{JavaScript Object Notation}
\acro{mitm}[MitM]{Man-in-the-Middle}
\acro{ml}[ML]{Machine Learning}
\acro{nids}[NIDS]{Network IDS}
\acro{obdii}[OBD\=/II]{on-board diagnostics II}
\acro{oem}[OEM]{Original Equipment Manufacturer}
\acro{ota}[OTA]{Over-the-air}
\acro{ot}[OT]{Operational Technology}
\acro{pcapng}[PCAPNG]{PCAP Next Generation}
\acro{per}[PER]{Packed Encoding Rule}
\acro{pii}[PII]{Personally Identifiable Information}
\acro{qsev}[QSEv]{Qualified Security Event}
\acro{sae}[SAE]{Society of Automotive Engineers}
\acro{sec}[SEC]{Security Event Center}
\acro{sem}[SEM]{Security Event Message}
\acro{sev}[SEv]{Security Event}
\acro{siem}[SIEM]{Security Information and Event Management}
\acro{soar}[SOAR]{Security Orchestration, Automation and Response}
\acro{soc}[SOC]{Security Operations Center}
\acro{sota}[SOTA]{Software-over-the-air}
\acro{tara}[TARA]{Threat Analysis and Risk Assessment}
\acro{ta}[TA]{Trust Anchor}
\acro{tcu}[TCU]{Telematic Control Unit}
\acro{tip}[TIP]{Threat Intelligence Platform}
\acro{v2i}[V2I]{Vehicle-to-Infrastructure communication}
\acro{v2v}[V2V]{Vehicle-to-Vehicle communication}
\acro{vids}[VIDS]{Vehicle Intrusion Detection System}
\acro{vsm}[VSM]{Vehicle Security Monitoring}
\acro{vsm}[VSM]{Vehicle Security Monitoring}
\acro{vsoc}[VSOC]{Vehicle Security Operations Center}
\acro{zkpok}[ZKPoK]{Zero Knowledge Proof of Knowledge}

\acro{ap}[AP]{Attacker Profile}
\acro{am}[AM]{Attack Motivation}
\acro{av}[AV]{Attack Vector}
\acro{at}[AT]{Attack Technique}
\acro{ati}[ATI]{Target and Impact}
\acroplural{ati}[ATIs]{Targets and Impacts}
\acro{ic}[IC]{Impact Class}
\acroplural{ic}[ICs]{Impact Classes}

\acro{mvb}[MVB]{Multifunction Vehicle Bus}
\acro{can}[CAN]{Controller Area Network}
\acro{soa}[SOA]{Service-oriented architecture}
\acro{lin}[LIN]{Local Interconnected Network}
\acro{uds}[UDS]{Unified Diagnostic Services}

\acro{tip}[TIP]{Threat Intelligence Platform}
\acro{ti}[TI]{Threat Intelligence}

\acro{snmp}[SNMP]{Simple Network Management Protocol}

\acro{s4l}[S4L]{Smart Sensor and SEC Security Level}
\acro{tee}[TEE]{Trusted Execution Environment}
\acro{hsm}[HSM]{Hardware Security Module}
\acro{tpm}[TPM]{Trusted Platform Module}

\acro{pki}[PKI]{Public Key Infrastructure}
\acro{csr}[CSR]{Certificate Signing Request}
\acro{ra}[RA]{Registration Authority}
\acro{ca}[CA]{Certification Authority}
\acro{ocsp}[OCSP]{Online Certificate Status Request}
\acro{crl}[CRL]{Certificate Revocation List}

 \acro{DSC}[DSC (0x10)]{DiagnosticSessionControl}
\acro{ER}[ER (0x11)]{ECUReset}
\acro{CDI}[CDI (0x14)]{ClearDiagnosticInformation}
\acro{RDI}[RDI (0x19)]{ReadDTCInformation}
\acro{RDBI}[RDBI (0x22)]{ReadDataByIdentifier}
\acro{RMBA}[RMBA (0x23)]{ReadMemoryByAddress}
\acro{RSDBI}[RSDBI (0x24)]{ReadScalingDataByIdentifier}
\acro{SA}[SA (0x27)]{SecurityAccess}
\acro{CC}[CC (0x28)]{CommunicationControl}
\acro{RDBPI}[RDBPI (0x2A)]{ReadDataByPeriodicIdentifier}
\acro{DDDID}[DDDID (0x2C)]{DynamicallyDefineDataIdentifier}
\acro{WDBI}[WDBI (0x2E)]{WriteDataByIdentifier}
\acro{IOCByID}[IOCByID (0x2F)]{InputOutputControlByIdentifier}
\acro{RC}[RC (0x31)]{RoutineControl}
\acro{RD}[RD (0x34)]{RequestDownload}
\acro{RU}[RU (0x35)]{RequestUpload}
\acro{TD}[TD (0x36)]{TransferData}
\acro{RTE}[RTE (0x37)]{RequestTransferExit}
\acro{RFT}[RFT (0x38)]{RequestFileTransfer}
\acro{WMBA}[WMBA (0x3D)]{WriteMemoryByAddress}
\acro{TP}[TP (0x3E)]{TesterPresent}
\acro{NR}[NR (0x7F)]{NegativeResponse}
\acro{ATP}[ATP (0x83)]{AccessTimingParameters}
\acro{SDT}[SDT (0x84)]{SecuredDataTransmission}
\acro{CDTCS}[CDTCS (0x85)]{ControlDTCSetting}
\acro{ROE}[ROE (0x86)]{ResponseOnEvent}
\acro{LC}[LC (0x87)]{LinkControl}
\acro{Auth29}[Auth29 (0x29)]{Authentication}

\acro{DID}[DID]{Data Identifier}
\acro{DTC}[DTC]{Diagnostic Trouble Code}
\acro{NRC}[NRC]{Negative Response Code}
\acro{PID}[PID]{Parameter Identifier}
\acro{RID}[RID]{Routine Identifier}
\acro{SID}[SID]{Service Identifier}

\acro{RE}[RE]{Reverse Engineering}
\end{acronym}
\newcommand{\acwithid}[1]{\ac{#1} (\acs{#1})}
\section{Introduction}
\label{sec:introduction}

The growing complexity and interconnectivity of modern vehicles have created notable security challenges. Vehicles are increasingly  susceptible to cyberattacks, which poses serious risks to both vehicle integrity and safety. This issue is tackled by the recent UN R155 regulation~\cite{UNR155}, which emphasizes the urgent requirement for strong cybersecurity management systems and protective measures. One essential layer of defense involves implementing effective security monitoring systems.

Cybersecurity challenges are particularly relevant for the \ac{uds} protocol~\cite{ISO14229}, which is the most commonly used diagnostic protocol in the automotive sector. UDS facilitates communication between vehicle \ac{ecu}s and diagnostic testers —- either external to the vehicle or vehicle-internal units. The services provided by UDS encompass a broad range of fundamental functionalities that the automotive industry utilizes throughout all phases of an ECU's lifecycle, including development, testing, operation, maintenance, and decommissioning. Consequently, these services are of significant interest to attackers, as they enable a high degree of control over the ECU. While the security of the UDS protocol has been explored in various studies~\cite{kulandaivel2024candid,lauser2023,matsubayashi2021attacks,weiss2021automated}, security monitoring for UDS has not been studied systematically before. 

This paper presents security monitoring strategies for the UDS protocol, wherein detection is based on in-vehicle logging and on processing log events in a remote \ac{vsoc}~\cite{mayer2024vehicle}. The \ac{vsoc} collects security events from the vehicle fleet and puts them in context with other data sources, e.g., vehicle records including maintenance plans, and threat intelligence digests. More concretely, firstly, we present log strategies specifying which security events are to be logged in vehicles. Secondly, we describe context data to be logged with security events. Finally, we describe detection strategies to analyze logged vehicle security events, with the goal to detect UDS attack scenarios. Strategies are formulated for the specific case of UDS but have the potential to generalize to the security monitoring of vehicle security events in general. 

We underline the relevance of the presented monitoring strategies by applying them to a
comprehensive taxonomy of UDS attack techniques~\cite{Yekta2025uds}. This taxonomy is based on Tactics, Techniques, and Procedures (TTP) and is structured along the automotive-specific `Vehicle Adversarial Tactics, Techniques, and Expert Knowledge' (VATT\&EK)~\cite{vattek1} and the more general MITRE `Adversarial Tactics, Techniques, and Common Knowledge' (ATT\&CK)~\cite{mitre_attck} frameworks. Our results show that presented detection strategies cover almost all of the attack techniques in this taxonomy. Among others. Moreover, our results show which attack techniques can be detected already on the vehicle side and which techniques require correlation of data sources in a fleet backend. We also show to which extent the security events standardized by a current industry standard, AUTOSAR, are already suited to support the detection of UDS attack techniques, and we identify corresponding gaps in the standard. 

In summary, we give an overview on detection strategies for attack techniques misusing the UDS protocol. In this way, our approach gives an example for developing security monitoring strategies for an automotive communication protocol. While \ac{vsoc} infrastructures have been established in recent years by vehicle manufacturers, it is still a challenge how to detect the occurrence of higher-level attack techniques based on low-level security events. The presented end-to-end monitoring strategies address this challenge. They can be used to implement UDS security monitoring, by deriving vehicle-side logging requirements and by guiding backend-side log processing in a VSOC. 

The paper is structured as follows. Section~{\ref{sec:related_work}} provides background and related work. Section~{\ref{sec:Methodology}} lays down the methodology used in this work and Section~{\ref{sec:evaluation}} presents the evaluation of the results. In Section~{\ref{sec:conclusion}}, we conclude our discussion and refer to possible future work.

\section{Background and Related Work}
\label{sec:related_work}
{\em Background.} Cybersecurity attacks have become a highly relevant threat for modern cars. First standards and regulations on security have already been created in the automotive industry. Examples are the ISO/SAE 21434 standard on vehicle cybersecurity~\cite{ISO21434} and the United Nations (UN)~R155~regulation~\cite{UNR155} providing cybersecurity provisions for vehicle type approval. The latter requests automotive manufacturers to be able to detect and respond to security attacks in their vehicles. For this goal, automotive manufacturers introduce security monitoring solutions for their vehicle fleets. 


{\em UDS.} In this work, we specifically consider security monitoring targeting to detect threat scenarios for the \ac{uds} protocol. The \ac{uds} protocol, standardized in~\cite{ISO14229}, is the most widely used protocol for vehicle diagnostics. It allows diagnostic tools to contact the \ac{ecu} installed in a vehicle which has \ac{uds} services enabled. Diagnostic services cover, among others, testing, calibration, or software updates. Table~\ref{tab:uds_services} provides an overview on \ac{uds} services. More details about the services can be found in~\cite{ISO14229}.

\begin{table}[!b]
\caption{UDS Services overview.}
\label{tab:uds_services}
\centering
\footnotesize
\setlength{\tabcolsep}{2pt}
\renewcommand{\arraystretch}{0.95} 

\begin{adjustbox}{max width=\linewidth}
\begin{tabular}{|p{1cm}|p{5.3cm}|p{2cm}|}
\hline
\textbf{SID} & \textbf{Service} & \textbf{Short} \\ \hline\hline
0x10 & DiagnosticSessionControl & DSC \\ \hline
0x11 & ECUReset & ER \\ \hline
0x14 & ClearDiagnosticInformation & CDTCI \\ \hline
0x19 & ReadDTCInformation & RDTCI \\ \hline
0x22 & ReadDataByIdentifier & RDBI \\ \hline
0x23 & ReadMemoryByAddress & RMBA \\ \hline
0x24 & ReadScalingDataByIdentifier & RSDBI \\ \hline
0x27 & SecurityAccess & SA \\ \hline
0x28 & CommunicationControl & CC \\ \hline
0x29 & Authentication & AUTH \\ \hline
0x2A & ReadDataByPeriodicIdentifier & RDBPI \\ \hline
0x2C & DynamicallyDefineDataIdentifier & DDDID \\ \hline
0x2E & WriteDataByIdentifier & WDBI \\ \hline
0x2F & InputOutputControlByIdentifier & IOCBI \\ \hline
0x31 & RoutineControl & RC \\ \hline
0x34 & RequestDownload & RD \\ \hline
0x35 & RequestUpload & RU \\ \hline
0x36 & TransferData & TD \\ \hline
0x37 & RequestTransferExit & RTE \\ \hline
0x38 & RequestFileTransfer & RFT \\ \hline
0x3D & WriteMemoryByAddress & WMBA \\ \hline
0x3E & TesterPresent & TP \\ \hline
0x83 & AccessTimingParameters & ATP \\ \hline
0x84 & SecuredDataTransmission & SDT \\ \hline
0x85 & ControlDTCSetting & CDTCS \\ \hline
0x86 & ResponseOnEvent & ROE \\ \hline
0x87 & LinkControl & LC \\ \hline
\end{tabular}
\end{adjustbox}
\end{table}





{\em UDS Security.} There are a number of reported vehicle vulnerabilities based on UDS services, e.g., \cite{CVE-2024-6348,Ronge2024KIASELTOS,pelechaty2024analysis}, which underlines the relevance to study security aspects of UDS. A focus of recent research on UDS security has been on implementation weaknesses of the UDS Security Access Service~\cite{durrwang2017security,van2018beneath,ring2014evaluation}. 
For first systematic evaluations of the attack surface of automotive diagnostic protocols, see~\cite{weiss2021automated,lauser2023}. 

A comprehensive analysis of attack techniques for UDS has been provided in~\cite{Yekta2025uds}. The derived taxonomy categorizes 53 UDS attack techniques along 9 tactics of known attack frameworks. Concretely, the used tactics are {\em Resource Development (RD)}, {\em Persistence (PS)}, {\em Privilege Escalation (PE)}, {\em Defense Evasion (DE)}, {\em Credential Access (CA)}, {\em Discovery (DS)}, {\em Lateral Movement (LM)}, {\em Collection (CL}), and {\em Affect Vehicle Function (AF)}. The attack techniques are used in the evaluation of detection strategies in Section~\ref{sec:evaluation} (Table~\ref{tab:uds_attacks_detection}).

{\em Security Monitoring.} As part of security monitoring solutions, in-vehicle software sensors are used to monitor automotive systems for security anomalies. Also research has so far focused on these on-board \acp{ids}, for an overview see~\cite{lampe2023intrusion}. \ac{nids} monitors in-vehicle networks, e.g., Controller Area Network (CAN) busses or Ethernet networks. \ac{hids} monitors in-vehicle electronic control units, e.g., on the operating system level or on their interfaces. The setup of \ac{vsoc}, i.e., backend infrastructures for fleet security monitoring, has been described in~\cite{LangerSchueppelStahlbock2019,grimm2024cyber,mayer2024vehicle}. 





{\em AUTOSAR Security Events.}
AUTOSAR is a firmware specification that is widely used in the automotive industry. AUTOSAR supports a set of Security Events (SEvs) for different technologies~\cite{FO-TR-SecurityEventsSpecification}, as well as modules to qualify SEvs~\cite{CP-SWS-IntrusionDetectionSystemManager} and distribute them on the network~\cite{FO-PRS-IntrusionDetectionSystem}. Within this work, we will compare our results with what has been standardized in AUTOSAR, to determine which functionality can be used off-the-shelf and where extensions are needed.

\section{Methodology}
\label{sec:Methodology}

In this section, a systematic strategy for UDS security monitoring is developed.
First, in Section \ref{sec:methodology:log-strat}, a set of logging strategies is defined that allows the generation of appropriate security-related logs in the vehicle components running UDS. Then, in Section~\ref{sec:methodology:context}, we provide a context data strategy, specifiying context data to be captured with vehicle security logs. Finally, in Section~\ref{sec:methodology:detection_strategies}, we define \textit{detection strategies} allowing to identify higher-level attack scenarios with high certainty. In general, detection can be executed both on the vehicle side as well as on the backend side in a \ac{vsoc}. However, in many cases, detection relies on the \ac{vsoc} receiving the data from the vehicle and validating it against information only available in offboard systems, in order to differentiate attacks from false positives.

\subsection{Logging Strategies}
\label{sec:methodology:log-strat}
This section defines the logging strategies that a vehicle and its subcomponents can implement to detect attacks on the \ac{uds} protocol. Due to constraints in the vehicle -- runtime, storage, connectivity limitations -- it is not possible to just record and send all data generated by the vehicle for analysis to a remote \ac{vsoc}. Therefore, we need to rely on an appropriate logging concept, defining which events are to be logged. In the following, we present a set of three logging strategies. 

\paragraph{Invalid Request (IR)} Logs are generated whenever a UDS request is recognized as invalid due to one of the following reasons.
\begin{itemize}
\item A UDS request is observed which does not satisfy input validation checks due to unexpected formats, parameters out of range, or invalid payloads.
\item A UDS request is observed under unexpected or non-permitted circumstances, at ECU or vehicle level, e.g., while the vehicle is driving at high speed or without required authorizations.
\end{itemize}

\paragraph{Function Execution (FE)} Log the execution of selected SIDs, due to their criticality for the security of the ECU. This can be used by \ac{vsoc} to validate if the operation makes sense in the context the vehicle is in. Examples are given by memory modifications or the execution of critical routines.

\paragraph{Message Flow Inconsistency (MFI)} Logs are generated whenever a UDS SID is recognized as inconsistently routed due to one of the following reasons.
\begin{itemize}
\item A message is observed with unexpected source.
\item A routed message is different from the original message.
\item A routed message appears without first seeing the original message.
\item Messages are observed in an unexpected sequence, e.g., multiple 0x27 seed requests are observed without a subsequent key response.
\end{itemize}
These logging strategies can then be activated or deactivated for each single UDS SID, according to the needs of the identified threats. Note that \textit{Invalid Request} and \textit{Function Execution} are both logging mechanisms that can be implemented by a \ac{hids} or a \ac{nids}, whereas implementing \textit{Message Flow Inconsistency} is more feasible as part of a \ac{nids}, since an overview of the different vehicle networks is needed.

\subsection{Log Context Data Strategy}
\label{sec:methodology:context}

Whenever one of the previously introduced logging strategies is activated, it generates a log. In order to enrich a log, to make it more useful for further analysis, it must be complemented with appropriate \textit{context data}. 

For the strategy \textit{Message Flow Inconsistency}, the context data strategy is always the same: the observed UDS SID, the targeted ECU, the observed request origin, and the expected request origin. 

For the strategies \textit{Invalid Request} and \textit{Function Execution}, context data depend on the associated UDS SID. Table \ref{tab:uds_log_context_data} specifies context data to be logged for these two logging strategies. 

\begin{table}[t]
\caption{Context data to be logged for each UDS Service with strategies Invalid Requests (IR) and Function Execution (FE).}
\label{tab:uds_log_context_data}
\centering
\scriptsize
\setlength{\tabcolsep}{2.6pt}
\renewcommand{\arraystretch}{0.92}

\begin{adjustbox}{max width=\linewidth}
\begin{tabular}{|p{0.9cm}|>{\raggedright\arraybackslash}p{6.7cm}|p{1.1cm}|}
\hline
\textbf{SID} & Context data to be logged for logging strategies \textbf{Invalid Requests} (1) and \textbf{Function Execution} (2) & \textbf{AR support} \\ \hline\hline
0x10 & SID\textsuperscript{(1, 2)}, SF\textsuperscript{(1, 2)}, NRC\textsuperscript{(1)} & - \\ \hline
0x11 & SID\textsuperscript{(1, 2)}, SF\textsuperscript{(1, 2)}, NRC\textsuperscript{(1)} & IR, FE \\ \hline
0x14 & SID\textsuperscript{(1, 2)}, groupOfDTC\textsuperscript{(1, 2)}, MemorySelection\textsuperscript{(1, 2)}, NRC\textsuperscript{(1)} & IR, FE \\ \hline
0x19 & SID\textsuperscript{(1, 2)}, SF\textsuperscript{(1, 2)}, NRC\textsuperscript{(1)} & - \\ \hline
0x22 & SID\textsuperscript{(1, 2)}, DID1\textsuperscript{(1, 2)}, ..., DIDn\textsuperscript{(1, 2)}, NRC\textsuperscript{(1)} & - \\ \hline
0x23 & SID\textsuperscript{(1, 2)}, memAddr\textsuperscript{(1, 2)}, memSize\textsuperscript{(1, 2)}, NRC\textsuperscript{(1)} & - \\ \hline
0x24 & SID\textsuperscript{(1, 2)}, DID\textsuperscript{(1, 2)}, NRC\textsuperscript{(1)} & - \\ \hline
0x27 & SID\textsuperscript{(1, 2)}, SF\textsuperscript{(1, 2)}, NRC\textsuperscript{(1)} & IR, FE \\ \hline
0x28 & SID\textsuperscript{(1, 2)}, SF\textsuperscript{(1, 2)}, NRC\textsuperscript{(1)} & IR, FE \\ \hline
0x29 & SID\textsuperscript{(1, 2)}, SF\textsuperscript{(1, 2)}, NRC\textsuperscript{(1)} & IR, FE \\ \hline
0x2A & SID\textsuperscript{(1, 2)}, transmissionMode\textsuperscript{(1, 2)}, periodicDID\#1\textsuperscript{(1, 2)}, ..., periodicDID\#n\textsuperscript{(1, 2)}, NRC\textsuperscript{(1)} & - \\ \hline
0x2C & SID\textsuperscript{(1, 2)}, SF\textsuperscript{(1, 2)}, dynamicallyDefinedDID\textsuperscript{(1, 2)}, sourceDID\#1\textsuperscript{(1, 2)}, ..., sourceDID\#n\textsuperscript{(1, 2)}, memAddr\textsuperscript{(1, 2)}, memSize\textsuperscript{(1, 2)}, NRC\textsuperscript{(1)} & - \\ \hline
0x2E & SID\textsuperscript{(1, 2)}, DID\textsuperscript{(1, 2)}, hash over dataRecord\textsuperscript{(1, 2)}, NRC\textsuperscript{(1)} & IR, FE \\ \hline
0x2F & SID\textsuperscript{(1, 2)}, DID\textsuperscript{(1, 2)}, I/O controlParameter\textsuperscript{(1, 2)}, NRC\textsuperscript{(1)} & IR, FE \\ \hline
0x31 & SID\textsuperscript{(1, 2)}, SF\textsuperscript{(1, 2)}, RID\textsuperscript{(1, 2)}, NRC\textsuperscript{(1)} & IR, FE \\ \hline
0x34 & SID\textsuperscript{(1, 2)}, memAddr\textsuperscript{(1, 2)}, memSize\textsuperscript{(1, 2)}, NRC\textsuperscript{(1)} & IR, FE \\ \hline
0x35 & SID\textsuperscript{(1, 2)}, memAddr\textsuperscript{(1, 2)}, memSize\textsuperscript{(1, 2)}, NRC\textsuperscript{(1)} & IR, FE \\ \hline
0x36 & SID\textsuperscript{(1)}, blockSequenceCounter\textsuperscript{(1)}, NRC\textsuperscript{(1)} & - \\ \hline
0x37 & SID\textsuperscript{(1, 2)}, NRC\textsuperscript{(1)}, hash over transferred data\textsuperscript{(2)} & - \\ \hline
0x38 & SID\textsuperscript{(1, 2)}, modeOfOperation\textsuperscript{(1, 2)}, filePathAndName\textsuperscript{(1, 2)}, NRC\textsuperscript{(1)} & IR, FE \\ \hline
0x3D & SID\textsuperscript{(1, 2)}, memAddr\textsuperscript{(1, 2)}, memSize\textsuperscript{(1, 2)}, NRC\textsuperscript{(1)}, hash over transferred data\textsuperscript{(2)} & IR, FE \\ \hline
0x3E & n/a & - \\ \hline
0x84 & SID\textsuperscript{(1, 2)}, Apar\textsuperscript{(1, 2)}, Signature/Encryption Calculation\textsuperscript{(1, 2)}, req. SID\textsuperscript{(1, 2)}, NRC\textsuperscript{(1)} & - \\ \hline
0x85 & SID\textsuperscript{(1, 2)}, SF\textsuperscript{(1, 2)}, NRC\textsuperscript{(1)} & IR, FE \\ \hline
0x86 & SID\textsuperscript{(1, 2)}, SF\textsuperscript{(1, 2)}, SID for response\textsuperscript{(1, 2)}, NRC\textsuperscript{(1)} & - \\ \hline
0x87 & SID\textsuperscript{(1, 2)}, SF\textsuperscript{(1, 2)}, NRC\textsuperscript{(1)} & -  \\ \hline
\multicolumn{3}{l}{} \\ 
\multicolumn{3}{l}{SID = service ID, SF = subfunction, NRC = negative response code,} \\ 
\multicolumn{3}{l}{DID = data identifier, other context data fields refer to parameters} \\ 
\multicolumn{3}{l}{defined in~\cite{ISO14229}.} \\

\end{tabular}
\end{adjustbox}
\end{table}

The column \textit{AR support} indicates whether AUTOSAR already provides security events for this UDS service, based on the logging strategies outlined before. The AUTOSAR security events define context data that is very well aligned with the proposal from Table~\ref{tab:uds_log_context_data}. The only differences are that AUTOSAR does not provide hashes over data for SIDs \ac{WDBI} and \ac{WMBA}, but it does provide the logical client source address for all UDS security events.

The proposed context data from Table~\ref{tab:uds_log_context_data}combines data from the UDS request and response and provides hence the security-relevant information in a compact form. Using the raw UDS requests/responses as context data is not recommended due to (1) possibly large messages (up to several hundred bytes e.g. for {\ac{Auth29}}, {\ac{TD}} or {\ac{WMBA}}) which could exhaust the resources of deeply embedded ECUs, (2) risk of information disclosure when sending UDS payload data in clear text to the VSOC and (3) separate SEvs for UDS requests and responses, which would need to be mapped in the VSOC and would prohibit the configuration of IR SEvs without FE SEvs.

Note that the presented logging strategies together with the context data strategy described in this subsection can generate a lot of false positives if applied indiscriminately, e.g., when activating \textit{Function Execution} for \ac{RDBI} without any additional conditions. Therefore, on top of the logging strategies, additional detection strategies must be defined, to differentiate between true attacks and false positives.

\subsection{Detection Strategies}
\label{sec:methodology:detection_strategies}

This section defines \textit{detection strategies} allowing to identify higher-level attack scenarios. Detection strategies are needed for two reasons. Firstly, many of the logs proposed in Section \ref{sec:methodology:log-strat} will also be generated under regular vehicle operations. Advanced checks and validations are needed to avoid false positive alerts. Secondly, there are attack scenarios which cannot be detected by vehicle-side logs alone. We introduce three detection strategies as follows. 


\paragraph{Suspicious Log Patterns (SLP)} 
\label{sec:methodology:detection_strategies:invalid_operation}
This detection strategy monitors for the occurrence of suspicious patterns in logs collected in the vehicle. They refer to failed, rejected or inconsistent UDS operations in the vehicle. This strategy includes pattern matching rules with counting. Counting is required to implement checks against thresholds, since, during regular vehicle operation, occasional failed UDS operations are to be expected. Therefore, for each SID service, a threshold defines how many failed operations are to be observed within a time interval before an alert is triggered. Detection of this category can be executed on the vehicle side.

\paragraph{Contextualized Log Checks (CLC)} 
\label{sec:methodology:detection_strategies:operations_consistency}
This detection strategy assesses the (successful or failed) execution of UDS services in context of additional information. Context information includes the vehicle state, vehicle records with maintenance and service plans, as well as summaries of preceding and succeeding logs. Vehicle records are usually maintained in a backend but not in a single vehicle. Concrete checks to be executed as part of this strategy are given as follows:
\begin{itemize}
\item Service calls are inconsistent with the vehicle status, e.g., workshop session, development/production mode.
\item Service call uses unexpected permissions.
\item Service call is inconsistent with vehicle configurations.
\item Service call is inconsistent with other logs, also from backend systems. 
\item In a service call, memory hashes do not match hashes of authentic software releases.
\item In a service call, DIDs or memory ranges rated as sensitive are referenced, e.g., when files or memory are to be read out or modified.
\end{itemize}
Detection of this category can be executed on the vehicle side only if required context data is available, otherwise it needs to be done in the backend.

\paragraph{Product Threat Intelligence (PTI)}
\label{sec:methodology:detection_strategies:product_intel}
This detection strategy uses threat intelligence information about the vehicle and its components to identify attack patterns. Sources for this can span from publicly available information, e.g., entries in public vulnerability databases, forums, or research papers, to confidentially disclosed information. Examples for the latter are supplier vulnerability disclosures, responsible vulnerability disclosures by white-hat-hackers, or internal penetration tests. For concrete cases, tags can be defined, including vehicle model, ECU type and attack patterns, to filter information feeds and to link them to concrete attack techniques.   
Alerts are then triggered whenever, based on this filtering, relevant information is identified.

In the implementation of detection strategies a)-c), a baseline of rules and their configuration is initially derived from the service specification of a vehicle model, and is finetuned based on evaluating false positive logs collected from test vehicles.

\begin{table*}[htp!]
\centering
\caption{UDS Attack Techniques and their Detection Strategies.}
\label{tab:uds_attacks_detection}
\footnotesize
    \begin{tabular}{|l|l|l|p{1.5cm}|p{1.5cm}|p{2cm}|}
    \hline
        \textbf{Attack ID} & \textbf{Attack Name} & \textbf{SIDs} & \textbf{Logging Strategies} & \textbf{AUTOSAR Support} & \textbf{Detection Strategies} \\ \hline
        AT-RD-1 & Firmware Reverse-Engineering & - & NA & No & PTI \\ \hline
        AT-RD-2 & Leak Secrets & ~ & NA & No & PTI \\ \hline
        AT-PS-1 & Download Custom Package & 0x34, 0x36, 0x37 & IR, FE & Only 0x34 & SLP, CLC, PTI \\ \hline
        AT-PE-1 & Change to Privileged Session & 0x10 & FE, MFI & No & CLC \\ \hline
        AT-PE-2 & Valid Credentials & 0x27, 0x29 & FE & \Checkmark & CLC, PTI \\ \hline
        AT-PE-3 & Replay Attack SA & 0x27 & IR, FE, MFI & \Checkmark & SLP, CLC, PTI \\ \hline
        AT-PE-4 & Brute-Force SA & 0x27 & IR, FE & \Checkmark & SLP, CLC \\ \hline
        AT-PE-5 & Weak Auth29 configurations & 0x29 & IR, FE & \Checkmark & CLC \\ \hline
        AT-DE-1 & Block DTCs Generation & 0x85 & FE & \Checkmark & CLC \\ \hline
        AT-DE-2 & Remove Attack Traces in DTCs & 0x14 & FE & \Checkmark & CLC \\ \hline
        AT-DE-3 & Replay Download & 0x34, 0x36, 0x37 & FE & Only 0x34 & CLC \\ \hline
        AT-DE-4 & Bypass Checks & Multiple & Various & No & CLC, PTI \\ \hline
        AT-DE-5 & Bypass Read Protections using DDDID & 0x2C, 0x22 & FE & No & CLC, PTI \\ \hline
        AT-CA-1 & Extract Secrets & 0x22, 0x23, 0x31 & FE & Only 0x31 & CLC \\ \hline
        AT-DS-1 & Service Discovery & Multiple & IR, FE & ( \Checkmark ) & SLP, CLC \\ \hline
        AT-DS-2 & Subfunction Discovery & Multiple & IR, FE & ( \Checkmark ) & SLP, CLC \\ \hline
        AT-DS-3 & Diagnostic Sessions Discovery & 0x10 & IR, FE & No & SLP, CLC \\ \hline
        AT-DS-4 & UDS Fuzzing & Multiple & IR, FE & ( \Checkmark ) & SLP, CLC \\ \hline
        AT-DS-5 & Check seed entropy in SA & 0x27 & IR, MFI & No & SLP \\ \hline
        AT-DS-6 & Reverse-engineer SA algorithm & 0x27 & FE & \Checkmark & CLC, PTI \\ \hline
        AT-DS-7 & Identify Auth29 configuration & 0x29 & FE & No & CLC, PTI \\ \hline
        AT-DS-8 & Enumerate algorithms, Auth29 & 0x29 & FE & No & CLC, PTI \\ \hline
        AT-DS-9 & Check challenge entropy, Auth29 & 0x29 & IR, MFI & No & SLP \\ \hline
        AT-DS-10 & Identify Configurations for SDT & 0x84 & FE & No & CLC, PTI \\ \hline
        AT-DS-11 & DID Enumeration & 0x22 & IR, FE & No & CLC, SLP \\ \hline
        AT-DS-12 & Routine Enumeration & 0x31 & IR, FE & \Checkmark & CLC, SLP \\ \hline
        AT-DS-13 & File System Discovery & 0x38 & IR, FE & \Checkmark & CLC, SLP \\ \hline
        AT-DS-14 & Eavesdropping & Multiple & NA & No & NA \\ \hline
        AT-LM-1 & Man-in-the-Middle & Multiple & IR, FE, MFI & ( \Checkmark ) & SLP, CLC, PTI \\ \hline
        AT-CL-1 & Event-Based Data Extraction & 0x86 & IR, FE & No & SLP, CLC \\ \hline
        AT-CL-2 & Periodic Data Extraction & 0x2A & IR, FE & No & SLP, CLC \\ \hline
        AT-CL-3 & DID Data Extraction & 0x22 & IR, FE & No & CLC \\ \hline
        AT-CL-4 & Memory Extraction & 0x23, 0x35 & IR, FE & Only 0x35 & CLC \\ \hline
        AT-CL-5 & File Extraction & 0x38 & IR, FE & \Checkmark & CLC \\ \hline
        AT-CL-6 & Read DTCs & 0x19 & IR, FE & No & SLP, CLC \\ \hline
        AT-AF-1 & Request Flooding & Multiple & IR, FE & ( \Checkmark ) & SLP, CLC, PTI \\ \hline
        AT-AF-2 & Request Blocking & Multiple & IR, FE, MFI & ( \Checkmark ) & PTI, SLP \\ \hline
        AT-AF-3 & Interrupt Operations, DSC & 0x10 & IR, FE, MFI & No & SLP \\ \hline
        AT-AF-4 & Impede Usage of SA & 0x27 & IR & \Checkmark & SLP \\ \hline
        AT-AF-5.1 & Resource Overload via ROE & 0x86 & IR, FE & No & SLP, CLC \\ \hline
        AT-AF-5.2 & Resource Overload via RDBPI & 0x2A & IR, FE, MFI & No & SLP, CLC \\ \hline
        AT-AF-6 & Interrupt Periodic Data Readout & 0x2A & IR, FE & No & SLP, CLC \\ \hline
        AT-AF-7 & Change IO Configuration & 0x2F & IR, FE, MFI & \Checkmark & SLP, CLC \\ \hline
        AT-AF-8 & Routine Misuse & 0x31 & FE & \Checkmark & CLC \\ \hline
        AT-AF-9 & Early Transfer Termination & 0x37 & IR, FE, MFI & No & SLP, CLC \\ \hline
        AT-AF-10 & Interrupt Routine & 0x31 & IR, FE, MFI & \Checkmark & SLP, CLC \\ \hline
        AT-AF-11 & Keep Session Open & 0x10, 0x3E & FE, MFI & No & CLC \\ \hline
        AT-AF-12 & I/O Control & 0x2F & IR, FE & \Checkmark & CLC \\ \hline
        AT-AF-13 & Disrupt ECU Communication & 0x28 & IR, FE, MFI & \Checkmark & CLC \\ \hline
        AT-AF-14 & Reset ECU & 0x11 & IR, FE, MFI & \Checkmark & SLP, CLC \\ \hline
        AT-AF-15 & DID Manipulation & 0x2E & IR, FE & \Checkmark & SLP, CLC \\ \hline
        AT-AF-16 & File Manipulation & 0x38 & IR, FE & \Checkmark & SLP, CLC \\ \hline
        AT-AF-17 & Memory Manipulation & 0x3D, 0x34 & IR, FE & \Checkmark & SLP, CLC \\ \hline
        \multicolumn{6}{p{.8\textwidth}}{--- Attack IDs refer to UDS attack techniques derived in~\cite{Yekta2025uds}, where IDs have the format AT-<TT>-<NO> where <TT> refers to the attack tactic and <NO> to the number of the attack technique in the respective category.}\\
        \multicolumn{6}{l}{--- (\Checkmark) refers to logging for supported SIDs only.}
    \end{tabular}

\end{table*}

\section{Evaluation}
\label{sec:evaluation}
This section evaluates the effectiveness of the logging and detection strategies presented in Section~\ref{sec:Methodology}. To this end, we applied the logging and detection strategies to a comprehensive taxonomy of attack techniques~\cite{Yekta2025uds}. For each attack technique of this taxonomy, we evaluated which strategies can be applied to detect the respective attack technique. The resulting mapping table is presented in Table~\ref{tab:uds_attacks_detection}. The table lists all attack techniques of this taxonomy, with their ID, name and affected UDS SIDs. Attack techniques are grouped by attack tactics. Columns "Logging Strategies" and "Detection Strategies" specify which strategies from Section~\ref{sec:Methodology} can be used to detect an occurrence of the respective attack technique. Moreover, column "AUTOSAR support" indicates that logging requirements of strategies IR and FE are already covered by the current AUTOSAR standardization.

Our evaluation focuses on three major topics. In Section~\ref{sec:evaluation:AUTOSARcoverage}, we focus on the logging aspects and compare our proposed logging strategies with the AUTOSAR-provided security events to identify gaps that need to be addressed in implementation projects. In Section~\ref{sec:evaluation:detection}, we discuss how to actually detect UDS attacks based on illustrative examples. Finally, in Section~\ref{sec:evaluation:takeaways} we draw conclusions and formulate take-away messages based on our analysis.

\subsection{AUTOSAR Logging Coverage}
\label{sec:evaluation:AUTOSARcoverage}

Efficient intrusion detection is relying on standardized logging strategies that are available off-the-shelf and hence easy to deploy and use. AUTOSAR lends itself as a basis for such an approach, due to its good acceptance in the automotive domain and native support for security events.

As shown in Table~\ref{tab:uds_log_context_data} and discussed in Section~\ref{sec:methodology:context}, AUTOSAR defines Security Events for 50\%  of the UDS services (13 of 26). The coverage analysis for the UDS attacks shown in Table~\ref{tab:uds_attacks_detection} is a bit more complex, since AUTOSAR does not provide support for all SIDs and can hence log certain attacks only partially. Out of the 53 attacks, AUTOSAR supports full logging for 20 and partial logging for an additional 10 attacks, rendering the overall logging support to 38-56\%.

While AUTOSAR provides a good basis for UDS attack logging, it fails at providing complete coverage. It is hence advised to introduce additional security events based on the context data proposal in Table~\ref{tab:uds_log_context_data}. This can be done by automotive manufacturers for their respective products, or directly in AUTOSAR by extending the available Security Events.

In addition, please note that AUTOSAR supports only the logging strategies IR and FE. MFI is not supported by AUTOSAR, since it is typically implemented as part of an \ac{nids}. Automotive manufacturers should take care that their \ac{nids} specification supports the MFI security event proposed in Section~\ref{sec:methodology:context}.

\subsection{UDS attack detection - examples}
\label{sec:evaluation:detection}

\begin{figure}
    \centering
    \includegraphics[width=\linewidth]{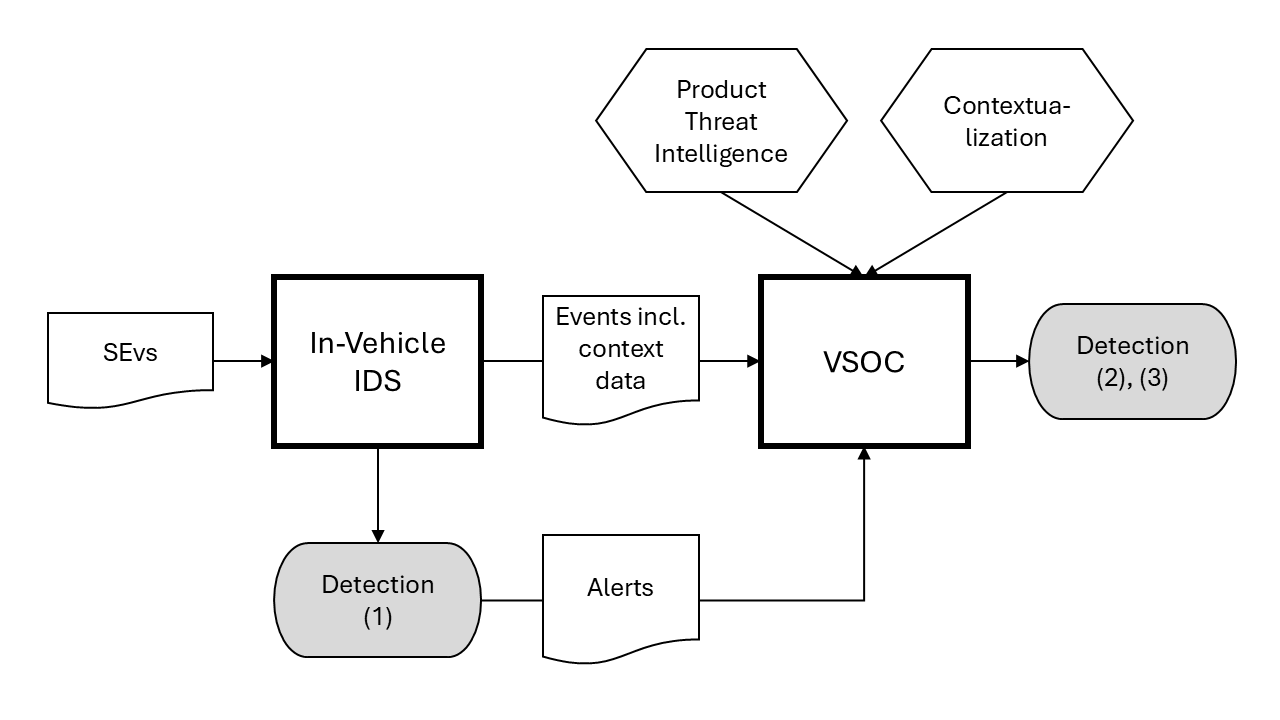}
    \caption{Detection process, including in-vehicle detection and VSOC-based detection. The numbers 
    refer to the examples from Section~{\ref{sec:evaluation:detection}}}.
    \label{fig:DefenderFlowchart}
\end{figure}

Detection of UDS attacks is very individual and strongly depending on the actual attack technique. Space restrictions do not allow to describe detection for every attack technique in detail. Instead, we illustrate the detection capabilities of our approach through three representative attack techniques, each demonstrating different aspects of our multi-layered security monitoring approach. Figure~{\ref{fig:DefenderFlowchart}} shows the general detection process, highlighting the detection possibilities in the vehicle and in the VSOC,
while locating the detection of the following examples.

\textbf{(1) AT-PE-4 Brute-Force SA Attack:} In this attack technique, an attacker tries to brute-force all possible response ("key") values for SA (0x27).
Applying logging strategy IR, Security Access brute-force attacks can be detected using existing AUTOSAR security events for SID 0x27, namely this is AUTOSAR security event 103 (SEV\_UDS\_SECURITY\_ACCESS\_
FAILED)~\cite{FO-TR-SecurityEventsSpecification}. By application of detection strategy SLP, multiple occurrences of this event within a short timeframe indicate a brute-force attempt against the Security Access service. This demonstrates effective detection using established AUTOSAR events with simple rate-based analysis. Additionally, detection strategy CLC may identify when authorizations are not consistent with the vehicle status.

\textbf{(2) AT-CL-3 DID Data Extraction:} In this attack technique, an attacker uses UDS service RDBI (0x22) to extract the information stored behind the DIDs, which may contain confidential data, e.g., keys. Detection of unauthorized RDBI operations is not possible through existing AUTOSAR security events, as no events are specified for this service. By application of logging strategies IR (logging unsuccessful access attempts) and FE (logging successful access), security events can address this gap by logging all accesses to sensitive DIDs, e.g., accessing cryptographic material. Context-data strategies ensure that DIDs are available as context data, and, for unsuccessful access attempts, the reason for rejection is available as Negative Response Code (NRC). Using strategy CLC, it can be ensured that only critical data identifier access attempts are captured, enabling detection of attacks targeting sensitive ECU information.

\textbf{(3) AT-PS-1 Download Custom Package:} In this attack technique, an attacker uses UDS services RD (0x34, request download), TD (0x36, transfer data), and RTE (0x37, request transfer exit) to download their own data into the ECU. Detection is possible by using logging strategies FE and IR, logging successful and unsuccessful invocation of relevant services (0x34, 0x36, 0x37). Context-data strategies ensure that firmware hashes are captured when completing download operations (0x37). By application of logging strategy CLC, these hashes are transmitted from the vehicle to the VSOC, where they are correlated with authorized firmware databases to detect downgrade attacks and unauthorized firmware installations. Logging strategies SLP and PTI can additionally be used to raise reliability of the detection, e.g., by detecting failed attempts in the operation, or by looking for known exploit patterns to install firmware. This attack cannot be detected by AUTOSAR security events alone, due to two fundamental limitations:
\begin{enumerate}
\item Attack detection requires firmware hash validation, which is not included in standard AUTOSAR security events.
\item Determining whether older or modified firmware is being installed requires backend knowledge of authorized firmware versions, which cannot be maintained locally in each vehicle.
\end{enumerate}

\subsection{UDS attack detection -  take-aways}
\label{sec:evaluation:takeaways}

Based on our analysis
from Section~\ref{sec:evaluation:detection}, we can compile the following take-away messages for detecting UDS attacks:


{\bf Vehicle-side detection can only cover a subset of UDS attack techniques.} Some attack techniques can be reliably detected on the vehicle-side. Examples are given by techniques of the Discovery tactic, e.g., service discovery or UDS Fuzzing, which can be detected by observing a large number of certain requests in a short time window. However, for the majority of attacks, the additional information and contextualization possibilities of a VSOC are needed for reliable detection, as described by the following two points.

{\bf Product Threat Intelligence is needed as part of a VSOC infrastructure.} For attack techniques of the attack tactic Resource Development, detection is possible using strategy PTI (Product Threat Intelligence) alone. Reverse engineering of firmware and leakage of secrets is usually done offline and can neither be detected by sensors in the vehicle nor by consistency analysis of logs in the backend. It can only be detected by observing reports of leakage of ECU firmware or UDS cryptographic material, e.g., in forums or news feeds. 

{\bf A {\em combination} of detection strategies as well as backend processing in a VSOC are needed for a maximum coverage and reliable detection of UDS attack techniques.} For many attack techniques, single detection strategies alone cannot provide sufficient evidence on the occurrence of an attack technique. However, the combination of detection strategies allows to reach a higher confidence by elimination of false positives. For example, consider AT-PE-1 "Change to Privileged Session" - an attacker using DSC (0x10) to change to a privileged session. In this case, the vehicle-side can log that DSC was called but needs additional data to distinguish whether this happened in context of a valid scenario, e.g., in context of a planned car service session.

\section{Conclusion and Future Work }
\label{sec:conclusion}

This paper presents multi-layered detection strategies for UDS-based attack techniques --- combining vehicle-level intrusion detection sensors with VSOC-level processing and threat intelligence. It is shown that strategies are suited to cover almost all elements from a comprehensive taxonomy of UDS techniques. 
Security monitoring strategies presented in this paper can be used as a guide to implement the detection of UDS attack techniques in a VSOC infrastructure: 

{\em Logging requirements.} Logging and context data strategies can be used as requirements for on-board intrusion detection components. The analysis from Table~\ref{tab:uds_attacks_detection} also shows in which cases we can refer to AUTOSAR standardized security events.

{\em Automated processing rules.} Detection strategies of the Suspicious Log Pattern and Contextualized Log Check categories can be used to define automated processing rules in a processing pipeline for aggregated onboard logs. Depending on system architecture, resources, and availability of context data, log processing may be done on the onboard side as well as on the backend side. Automated processing results in alerts to be handled in an incident management system.

{\em Threat intelligence triggers.} Detection strategies of the Product Threat Intelligence category can be used to define trigger criteria for the evaluation of threat intelligence information. Depending on the trigger critria, news feeds will be filtered down towards notifications relevant for the UDS monitoring use cases, and can be linked to alerts.  

{\em Playbooks.} On a higher level, detection scenarios can be implemented in playbooks, guiding the validation of alerts in an incident management system, also including manual analysis steps. Each UDS security attack technique can be covered by a playbook, while alerts with similar processing steps can be bundled in a joint playbook. 

In this way, this paper gives concrete guidelines on building VSOC detection scenarios based on the UDS protocol, and our accepted follow-up describes a VSOC for automotive and rail, specifying formats for vehicle security events and alerts, as well as detection and response capabilities \cite{Yekta2025Holistic}.

While this paper gives a qualitative assessment of detection strategies, their experimental evaluation with real vehicles remains a topic for future work.

\section*{Acknowledgment}
This research is accomplished within the project “FINESSE” (FKZ 16KIS1584K). We acknowledge the financial support for the project by the Federal Ministry of Research, Technology and Space (BMFTR).

\bibliographystyle{IEEEtran}
\bibliography{finesse-uds}

\end{document}